\documentclass[letterpaper, conference]{IEEEtran}
\usepackage{fullpage,graphicx,psfrag,url,setspace}
\usepackage{cite}
\usepackage{epsf}
\usepackage{color}
\usepackage[small,bf]{caption}
\usepackage{amsmath,verbatim,amssymb,amsfonts,amscd, graphicx,algorithmic}
\usepackage{hyperref}
 \hypersetup{colorlinks=true}
\newtheorem{thm}{Theorem}

\newtheorem{lemma}[thm]{Lemma}

\newcommand{\vect}{\boldsymbol} %alternative use \vec

\newcommand{\lbb}{\left\{}
\newcommand{\rbb}{\right\}}
\newcommand{\lsb}{\left(}
\newcommand{\rsb}{\right)}
\newcommand{\lmb}{\left[}
\newcommand{\rmb}{\right]}

\newcommand{\ben}{\begin{eqnarray}}

\newcommand{\een}{\end{eqnarray}}

\title{Diversity-Multiplexing-Delay Tradeoffs in MIMO Multihop Networks with ARQ}

\author{%
  \authorblockN{Yao Xie\authorrefmark{1},
    Andrea Goldsmith\authorrefmark{1}
  }\\
  \authorblockA{%
    \authorrefmark{1}Department of Electrical Engineering, Stanford University, Stanford, CA.\\
  }

  Email: yaoxie@stanford.edu, andrea@wsl.stanford.edu

}

\begin{document}
\maketitle

\begin{abstract} Tradeoff in diversity, multiplexing, and delay in
multihop MIMO relay networks with ARQ is studied, where the random
delay is caused by queueing and ARQ retransmission. This leads to
an optimal ARQ allocation problem  with per-hop delay or
end-to-end delay constraint. The optimal ARQ allocation has to
trade off between the ARQ error that the receiver fails to decode
in the allocated maximum ARQ rounds and the packet loss due to
queueing delay. These two probability of errors are characterized
using the diversity-multiplexing-delay tradeoff (DMDT) (without
queueing) and the tail probability of random delay derived using
large deviation techniques, respectively. Then the optimal ARQ
allocation problem can be formulated as a convex optimization
problem. We show that the optimal ARQ allocation should balance
each link performance as well avoid significant queue delay, which
is also demonstrated by numerical examples.
\end{abstract}

\section{Introduction}

%Another reference \cite{FallahiHossain2006}. \cite{Ganesh1998}.

\footnote{01/04/2010. Submitted to The IEEE International
Symposium on Information Theory 2010.}In a multihop relaying
system, each terminal receives the signal only from the previous
terminal in the route and, hence, the relays are used for coverage
extension. Multiple input-multiple output (MIMO) systems can
provide increased data rates by creating multiple parallel
channels and increasing diversity by robustness against channel
variations. Another degree of freedom can be introduced by an
automatic repeat request (ARQ) protocol for retransmissions. With
the multihop ARQ protocol, the receiver at each hop feeds back to
the transmitter a one-bit indicator on whether the message can be
decoded or not. In case of a failure the transmitter sends
additional parity bits until either successful reception or
message expiration. The ARQ protocol provides improved reliability
but also causes transmission delay of packets. Here we study a
multihop MIMO relay system using the ARQ protocol. Our goal is to
characterize the tradeoff in speed versus reliability for this
system.

The rate and reliability tradeoff for the point-to-point MIMO
system, captured by the diversity-multiplexing tradeoff (DMT), was
introduced in \cite{ZhengTse03diversity}. Considering delay as the
third dimension in this asymptotic analysis with infinite SNR, the
diversity-multiplexing-delay tradeoff (DMDT) analysis for a
point-to-point MIMO system with ARQ is studied in
\cite{GamalDamen06themimo}, and the DMDT curve is shown to be the
scaled version of the corresponding DMT curve without ARQ. The
DMDT in relay networks has received a lot of attention as well
(see, e.g., \cite{Tabet:IT:07}.) In our recent work
\cite{YaoDenizGoldsmith2009}, we extended the point-to-point DMDT
analysis to multihop MIMO systems with ARQ and proposed an ARQ
protocol that achieves the optimal DMDT.

The DMDT analysis assumes asymptotically infinite SNR. However, in
the more realistic scenario of finite SNR, retransmission is not a
negligible event and hence the queueing delay has to be brought
into the picture (see discussions in
\cite{HollidayGoldsmithPoor2008}). With finite SNR and queueing
delay, the DMDT will be different from that under the infinite SNR
assumption. The DMDT with queueing delay is studied in
\cite{HollidayGoldsmithPoor2008} and an optimal ARQ adapted to the
instantaneous queue state for the point-to-point MIMO system is
presented therein.

In this work, we extend the study \cite{HollidayGoldsmithPoor2008}
of optimal ARQ assuming high but finite SNR and queueing delay in
point-to-point MIMO systems to multihop MIMO networks. This work
is also an extention our previous results in
\cite{YaoDenizGoldsmith2009} to incorporate queueing delay. We use
the same metric as that used in \cite{HollidayGoldsmithPoor2008},
which captures the probability of error caused by both ARQ error,
and the packet loss due to queueing delay. The ARQ error is
characterized by information outage probability, which can be
found through a diversity-multiplexing-delay tradeoff analysis
\cite{GamalDamen06themimo,YaoDenizGoldsmith2009}. The packet loss
is given by the limiting probability of the event that packet
delay exceeds a deadline. Unlike the standard queuing models for
networks (e.g., \cite{BisnikAbouzeid2006,BolchGreinerdeMeer2006})
where only the number of messages awaiting transmission is
studied, here we also need to study the amount of time a message
has to wait in the queue of each node. Our approach is slightly
different from \cite{HollidayGoldsmithPoor2008}, where the optimal
ARQ decision is adapted per packet; we study the queues after they
enter the stable condition, and hence we use the stationary
probability of a packet missing a deadline. An immediate tradeoff
in the choice of ARQ round is: the larger the number of ARQ
attempts we used for a link, the higher the diversity and
multiplexing gain we can achieve, meaning a lower ARQ error.
However, this is at a price of more packet missing deadline. Our
goal is to find an optimal ARQ allocation that balances these two
conflicting goals and equalizes performance of each hop to
minimizes the probability of error.

The remainder of this paper is organized as follows. Section II
introduces system models and the ARQ protocol. Section III
presents our formulation and main results. Numerical examples are
shown in Section IV. Finally Section V concludes the paper.

\section{Models and Background}

\subsection{Channel and ARQ Protocol Models}

\begin{figure}[h]
    \begin{center}
        \includegraphics[width=2.6in]{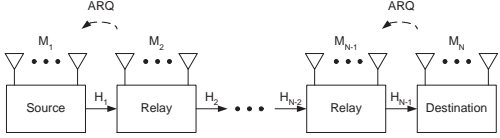}
    \end{center}
    \caption{Upper: relay network with direct link from source to
    destination. Lower: multihop MIMO
    relay network without direct link.}
    \label{Fig:models}
\end{figure}

Consider a multihop MIMO network consisting of $N$ nodes: with the
source corresponding to $i = 1$, the destination corresponding to
$i = N$, and $i = 2, \cdots, N-1$ corresponding to the
intermediate relays, as shown in Fig. \ref{Fig:models}. Each node
is equipped with $M_i$ antennas. The packets enter the network
from the source node, and exit from the destination node, forming
an open queue. The network uses a multihop automatic repeat
request (ARQ) protocol for retransmission. With the multihop ARQ
protocol, in each hop, the receiver feeds back to the transmitter
a one-bit indicator about whether the message can be decoded or
not. In case of a failure the transmitter retransmits. Each
channel block for the same message is called an ARQ round. We
consider the fixed ARQ allocation, where each link $i$ has a
maximum of ARQ rounds $L_i$, $i = 1, \cdots N-1$. The packet is
discarded once the maximum round has been reached. The total
number of ARQ rounds is limited to $L$: $\sum_{i=1}^{N-1}L_i \leq
L$. This fixed ARQ protocol has been studied in our recent paper
\cite{YaoDenizGoldsmith2009}.

\begin{figure}[h]
    \begin{center}
    \includegraphics[width=1.68in]{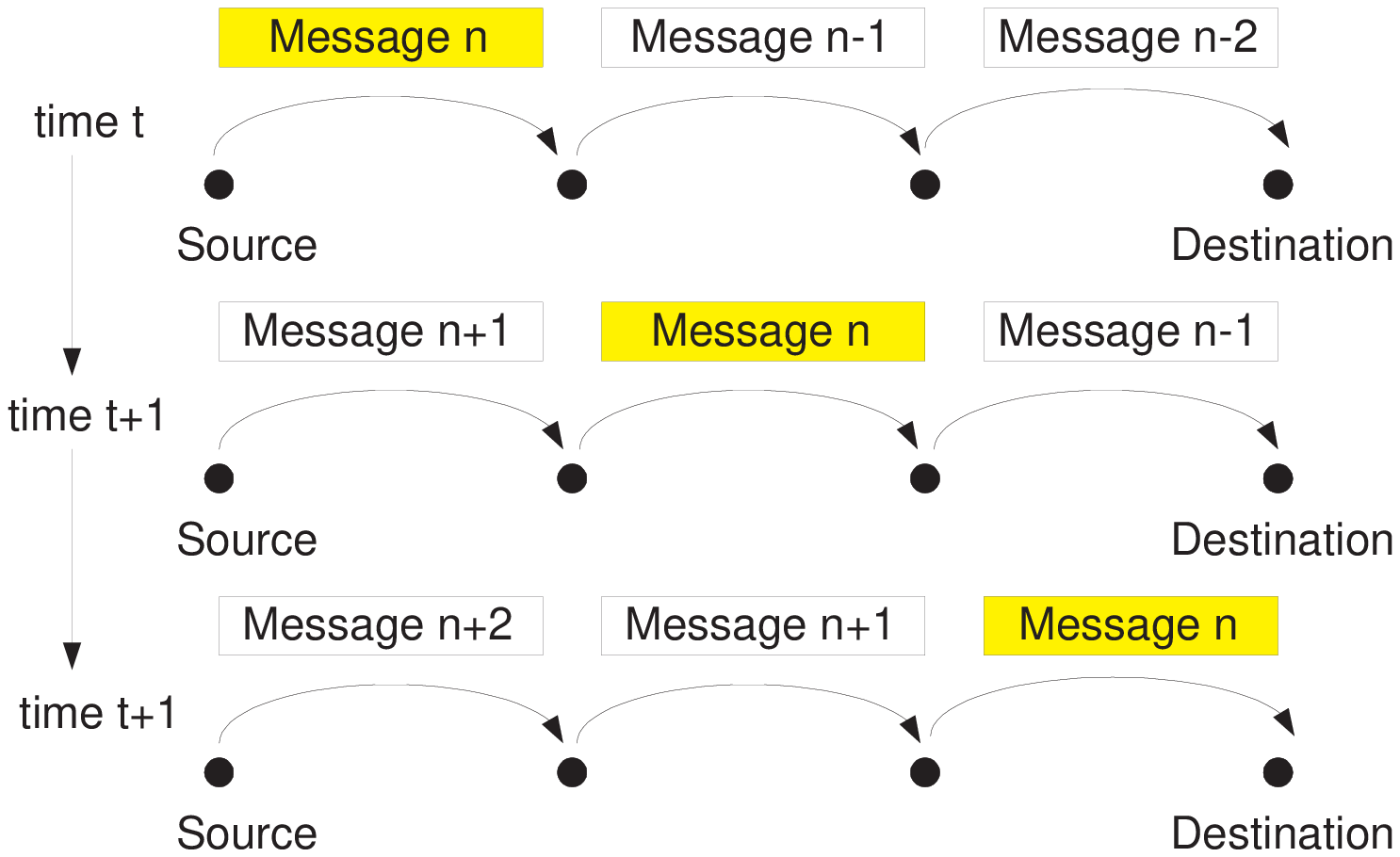}
        \includegraphics[width=1.68in]{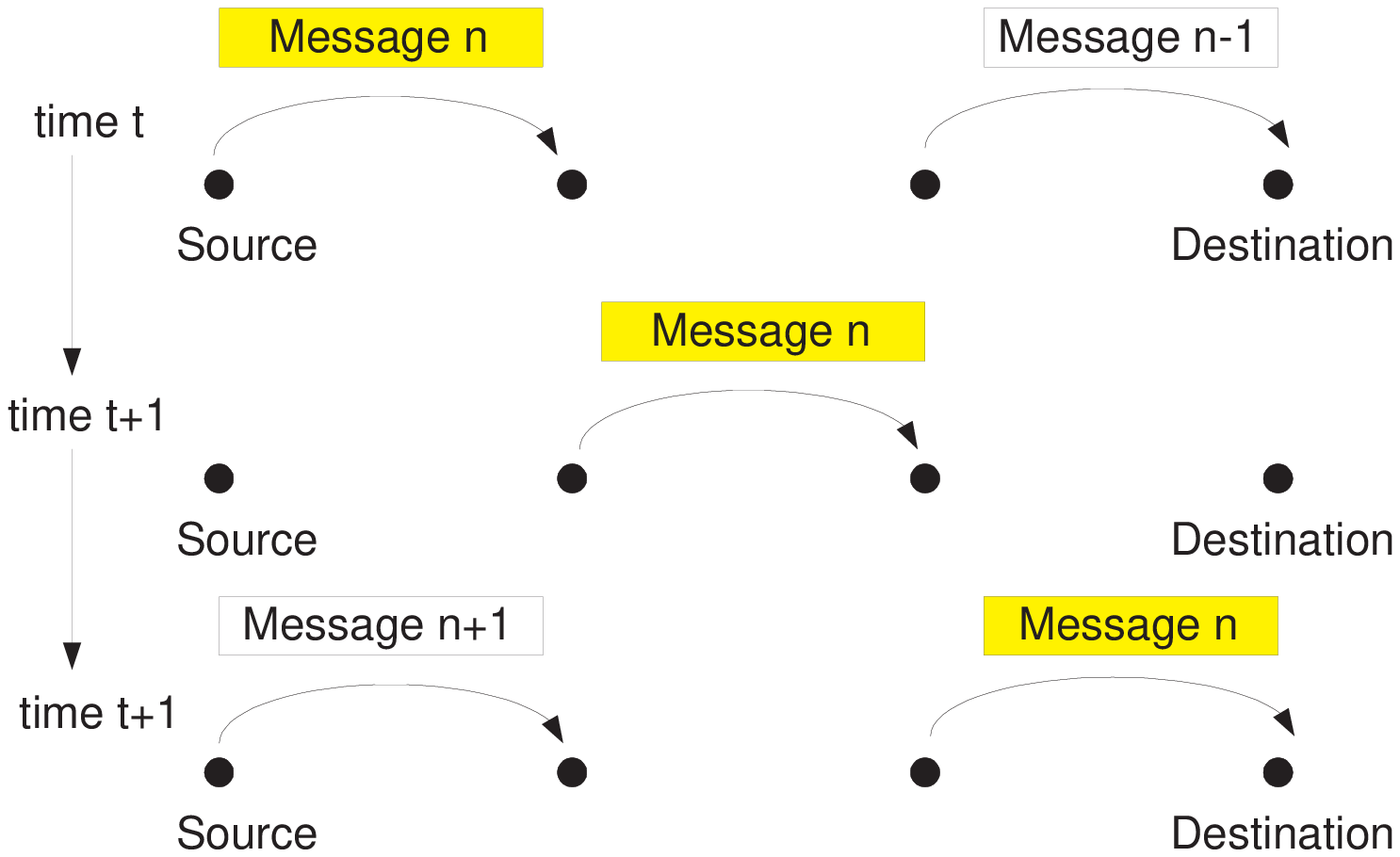} %\\(a) ~~~~~~~~~~~~~~~~~~~~~~~~~~~~~(b)
    \end{center}
    \caption{Left: full duplex multihop relay network. Right: half duplex relay
    multihop MIMO relay network.}
    \label{Fig:duplex}
\end{figure}

Assume the packets are delay sensitive: the end-to-end
transmission delay cannot exceed $k$. One strategy to achieve this
goal is to set a deadline $k_i$ for each link $i$ with $\sum_{i =
1}^{N-1} k_i \leq k$. Once a packet delays more than $k_i$ it is
removed from the queue. This per-hop delay constraint corresponds
to the finite buffer at each node. Another strategy is to allow
large per-hop delay while imposing an end-to-end delay constraint.
Other assumptions we have made for the channel models are
\begin{itemize}
\item[(i)] The channel between the $i$th and ($i+1$)th nodes is
given by:
\begin{eqnarray}
\vect{Y}_{i,l} = \sqrt{\frac{SNR}{M_i}} \vect{H}_{i,l}
\vect{X}_{i,l} + \vect{W}_{i,l}, \quad 1 \leq l \leq L_i.
\end{eqnarray}
The message is encoded by a space-time encoder into a sequence of
$L$ matrices $\{ \vect{X}_{i,l} \in \mathcal{C}^{M_i \times T}, :
l = 1, \cdots, L \}$, where $T$ is the block length, and
$\vect{Y}_{i,l} \in \mathcal{C}^{M_{i+1} \times T}$, $i = 1,
\cdots, N-1$, is the received signal at the $(i+1)$th node, in the
$l$th ARQ round. The rate of the space-time code is $R$. Channels
are assumed to be frequency non-selective, block Rayleigh fading
and independent of each other, i.e., the entries of the channel
matrices $\vect{H}_{i,l} \in \mathcal{C}^{M_{i+1}\times M_i}$ are
independent and identically distributed (i.i.d.) complex Gaussian
with zero mean and unit variance. The additive noise terms
$\vect{W}_{i,l}$ are also i.i.d. complex Gaussian with zero mean
and unit variance. The forward links and ARQ feedback links only
exist between neighboring nodes.
\item[(ii)] We consider both the full-duplex and half-duplex
relays (see, e.g., \cite{YaoDenizGoldsmith2009}) %\cite{LanemanTseWornell2003},
where the relays can or cannot transmit and receive at the same
time, respectively, as shown in Fig. \ref{Fig:duplex}. Assume the
relays use a decode-and-forward protocol
(see, e.g., \cite{YaoDenizGoldsmith2009}). %\cite{AzarianElGamal2005}.
\item[(iii)] We assume a short-term power constraint at each node
for each block code. Hence we do not consider power control.
\item[(iv)] We consider both the long-term static channel, where
$\vect{H}_{i,l} = \vect{H}_i$ for all $l$, i.e. the channel state
remains constant during all the ARQ rounds, and independent for
different $i$. Our results can be extended to the the short-term
static channel using the DMDT analysis given in
\cite{YaoDenizGoldsmith2009}.

%In the long-term static channel model, the channel state is
%assumed to be static within each block. However in the short-term
%static channel model, we assume that the channel state varies from
%one channel use to another independently. This provides additional
%time-diversity, hence, higher robustness against fading.
\end{itemize}

\begin{figure}[h]
    \begin{center}
        \includegraphics[width=1.7in]{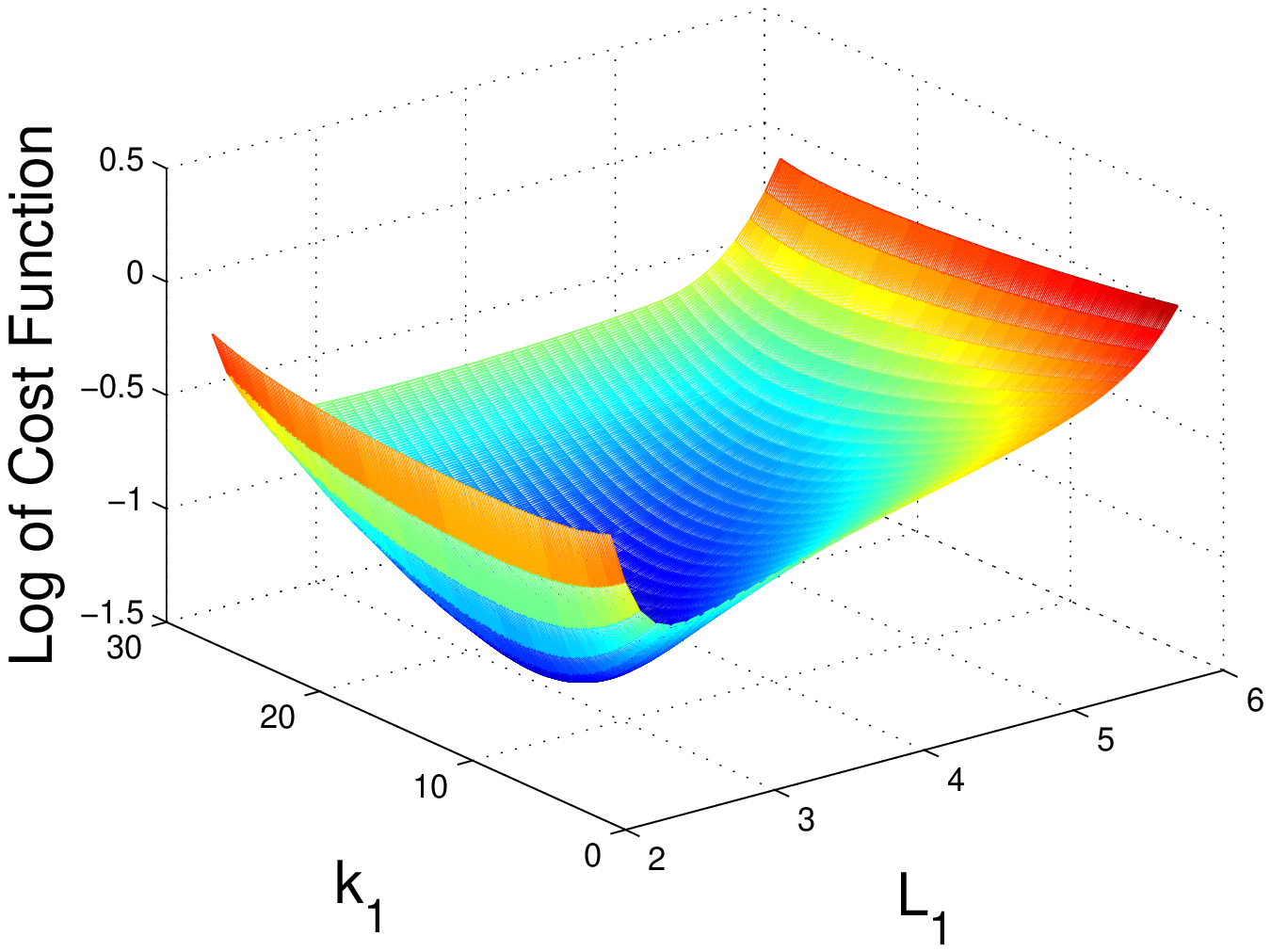}
    \end{center}
    \caption{The logarithm of the cost function (\ref{tran_dist_1})
    for the (4, 1, 2) multihop MIMO relay networks. SNR is 20 dB.}
    \label{Fig:3Deg}
\end{figure}

\begin{figure}[h]
    \begin{center}
        \includegraphics[width=1.5in]{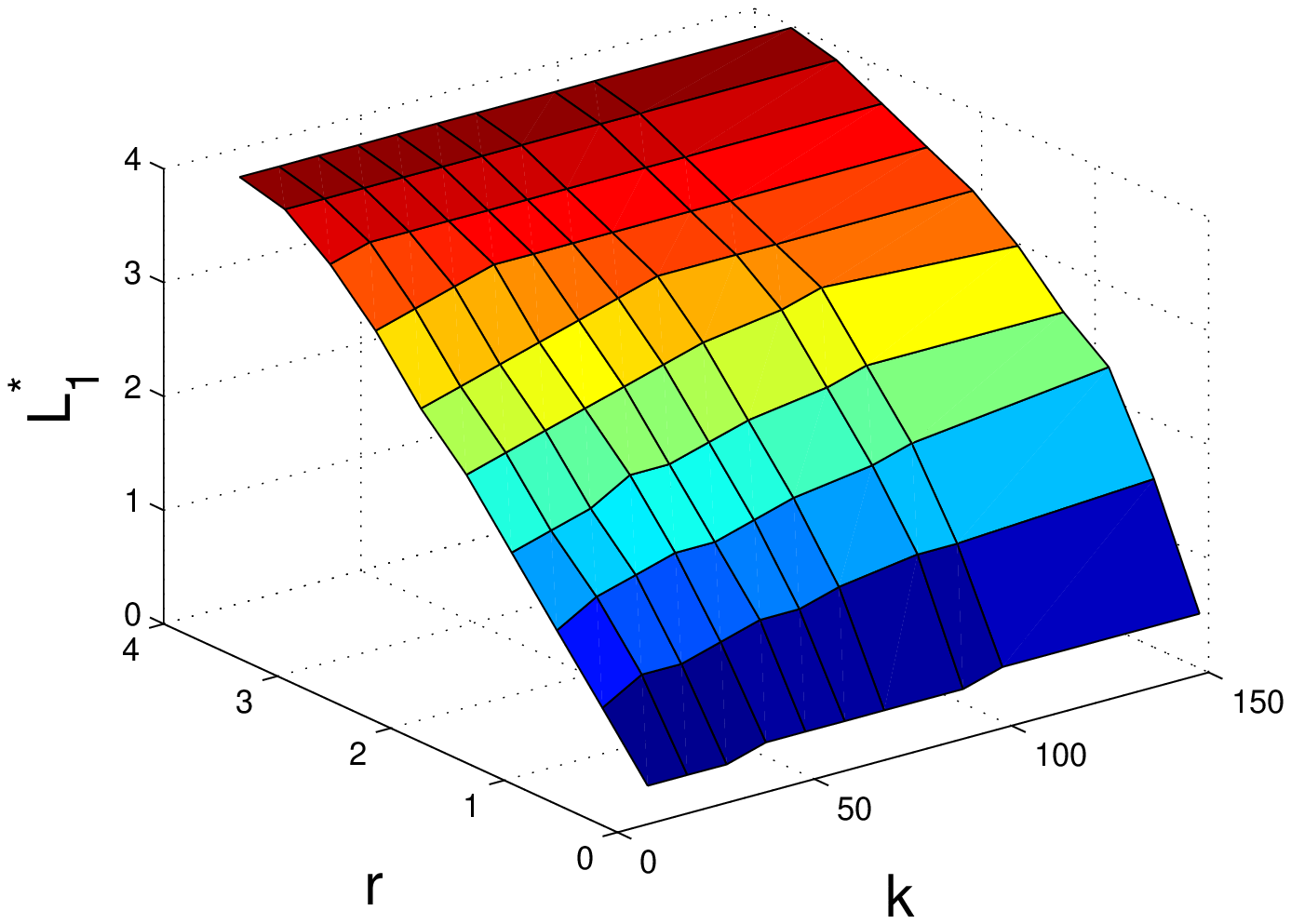}
        \includegraphics[width=1.5in]{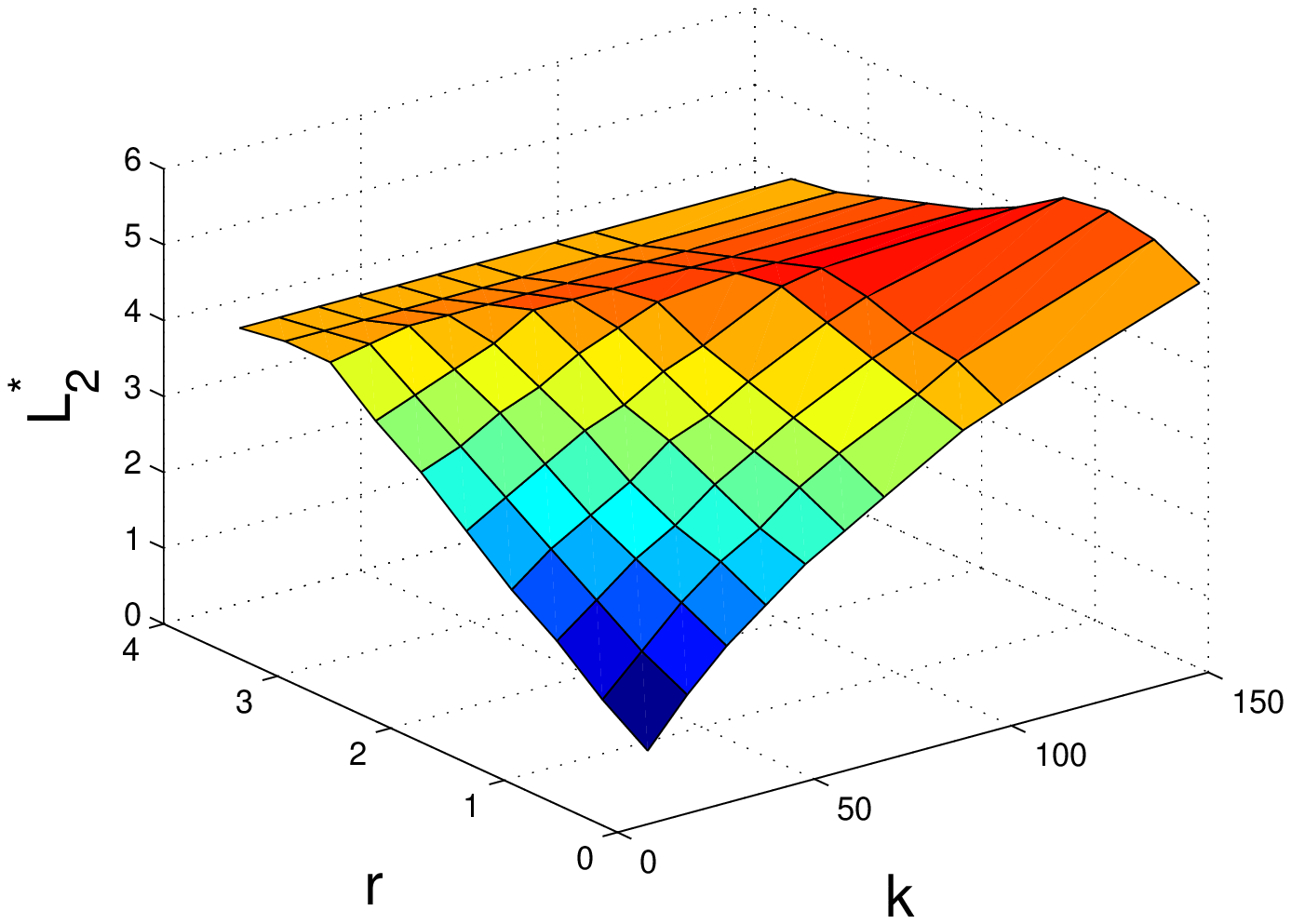}
        \includegraphics[width=1.5in]{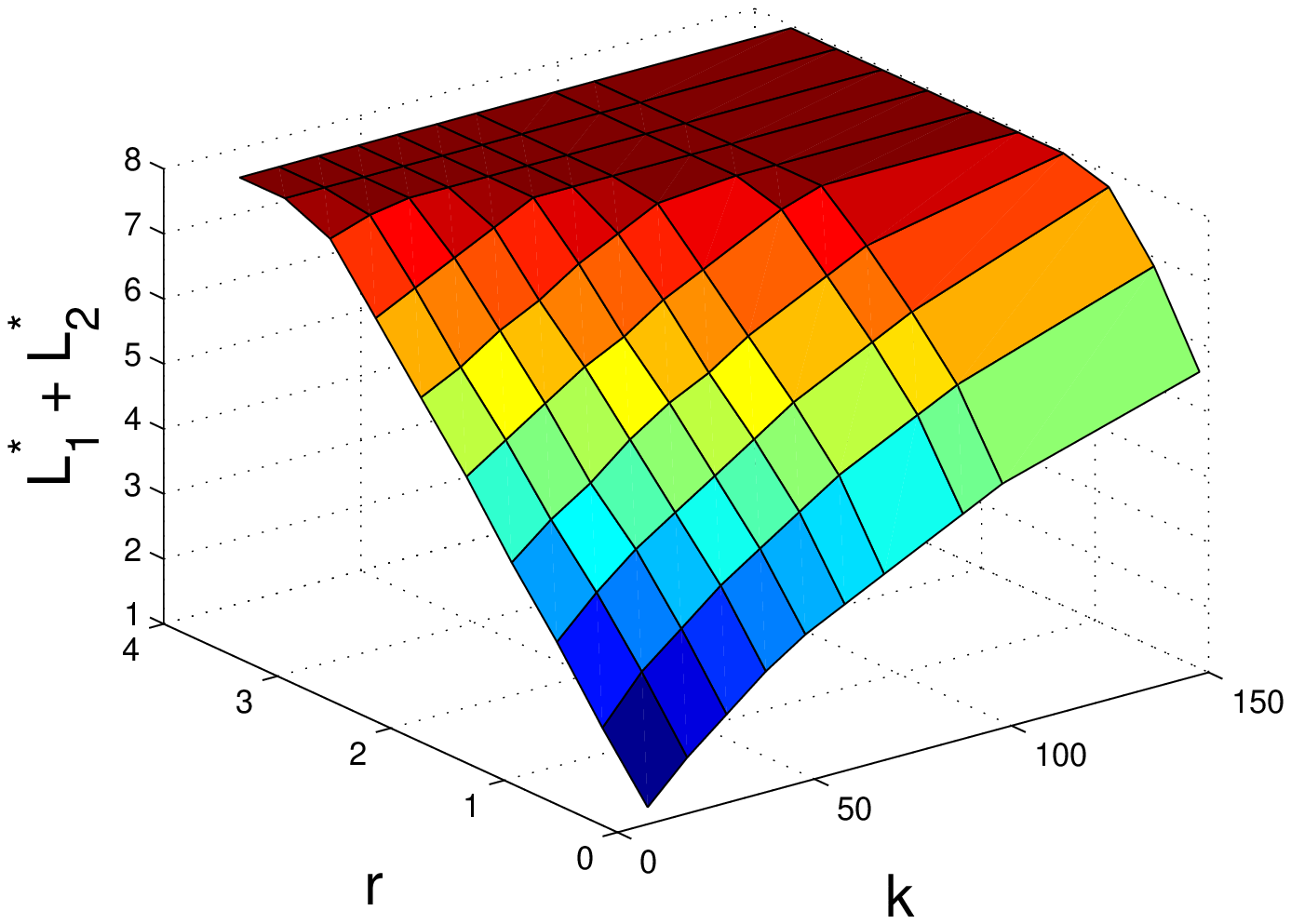}
        \includegraphics[width=1.5in]{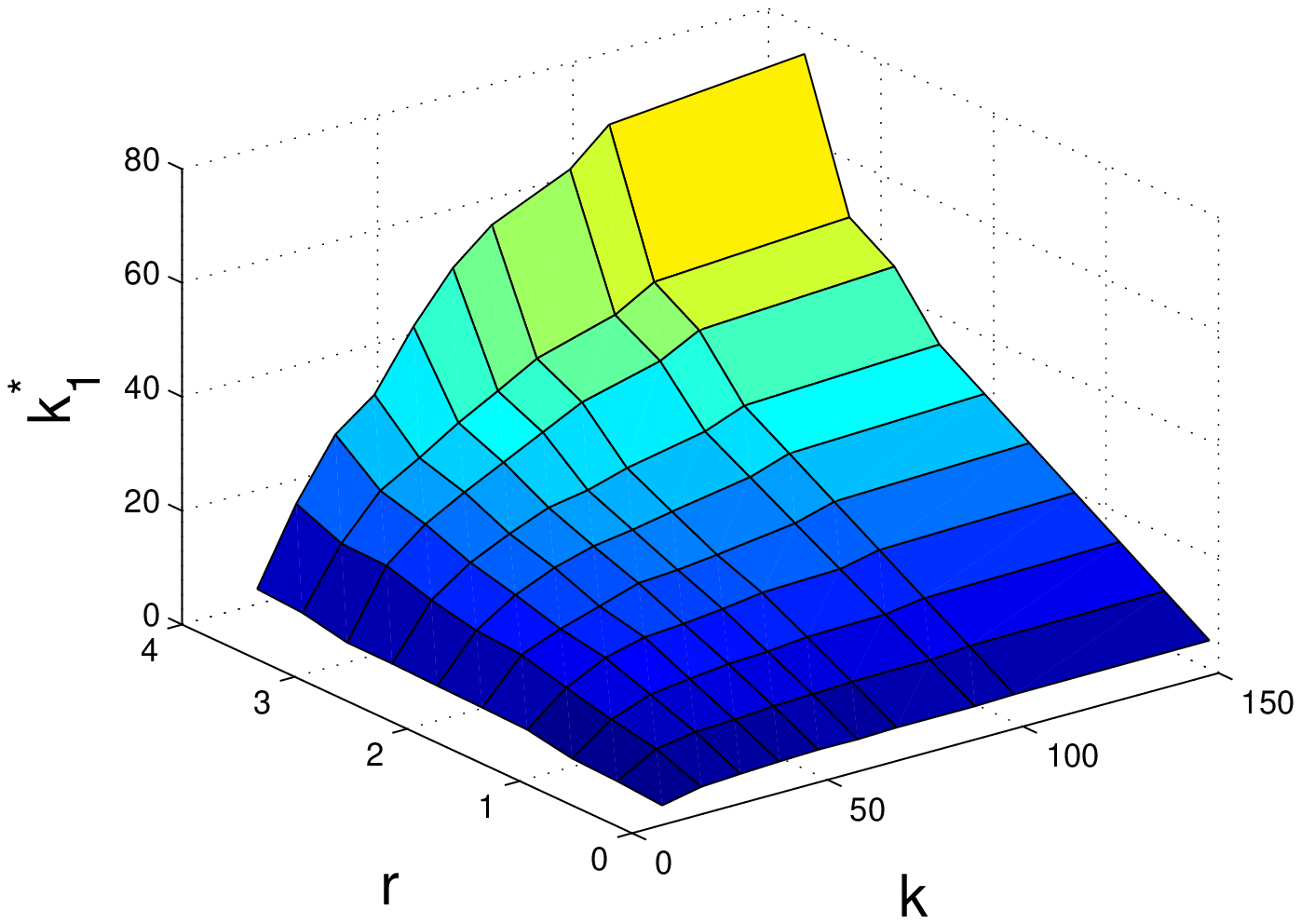}
    \end{center}
    \caption{Allocations of optimal ARQ: $L_1^*$, $L_2^*$, $L_1^* +
    L_2^*$, $k_1^*$, in a (4, 1, 2) MIMO relay network. SNR is 20 dB. (The optimal $k_2^* = k - k_1^*$.)}
    \label{Fig:threeHop422}
\end{figure}

\subsection{Queueing Network Model}

We use an $M/M/1$ queue tandem to model the multihop relay
networks. The packets arrive at the source as a Poisson process
with mean interarrival time $\mu$, (i.e., the time between the
arrival of the $n$th packet and $(n-1)$th packet.)
%Thus the average number of packets arrive per ARQ round is $1/\mu$.
The random service time depends on the channel state and is upper
bounded by the maximum ARQ rounds allocated $L_i$. As an
approximation we assume the random service time at Node $i$ for
each message is i.i.d. with exponential distribution and mean
$L_i$. With this assumption we can treat each node as an $M/M/1$
queue. This approximation makes the problem tractable and
characterizes the qualitative behavior of MIMO multihop relay
network. Node $i$ has a finite buffer size. The packets enter into
the buffer and are first-come-first-served (FCFS). Assume $\mu
\geq L_i$ so that the queues are stable, i.e., the waiting time at
a node does not go to infinity as time goes on.
%
%For the multihop relay network, when the relays are full duplex,
%there are $N-1$ buffers at the source and all the relay nodes
%because there is no direct link between source and destination and
%the relays can transmit and receive at the same time; when the
%relays are half duplex, there are $N-2$ buffers at the source and
%the first $N-3$ relays.
%
Burke's theorem (see, e.g., \cite{BolchGreinerdeMeer2006}) says
that the packets depart from the source and arrive at each relay
as a Poisson process with rate $p_i/\mu$, where $p_i$ is the
probability that a packet can reach the $i$th node. With high SNR,
the packet reaches the subsequent relays with high probability:
$p_i \approx 1$ (the probability of a packet dropping is small
because it uses up the maximum ARQ round.) Hence all nodes have
packets arrive as a Poisson process with mean inter-arrival time
$\mu$.

\subsection{Throughput}

Denote by $b$ the size of the information messages in bits, $B[t]$
the number of bits removed from transmission buffer at the source
at time slot $t$. Define a renewal event as the event that the
transmitted message leaves the source and eventually is received
by the destination node possibly after one or more ARQ
retransmissions. We assume that under full-duplex relays the
transmitter cannot send a new message until the previous message
has been decoded by the relay at which point the relay can begin
transmission over the next hop (Fig. \ref{Fig:duplex}a.) Under
half duplex relays we assume transmitter cannot send a new message
until the relay to the next hop completes its transmission (Fig.
\ref{Fig:duplex}b.)

The number of bits $\bar{B}$ transmitted in each renewal event,
for full-duplexing $\bar{B} = (N-1)b$, and for half-duplexing
$\bar{B} = (N-1)b/2$ when $N$ is odd, and $\bar{B} = Nb/2$ when
$N$ is even. The long-term average throughput of the ARQ protocol
is defined as the transmitted bits per channel use (PCU)
\cite{GamalDamen06themimo}, which can be found using renewal
theory \cite{Ross1995}:
\begin{eqnarray}
\eta &=& \liminf_{s\rightarrow \infty}\frac{1}{Ts} \sum_{t=1}^s
B[t] = \frac{\bar{B}}{E(\tau)}
\doteq \frac{\bar{B}}{(N-1)T} \nonumber \\
&=& \left\{%
\begin{array}{ll}
    R, & \hbox{Full duplex;} \\
    \frac{R}{2}, & \hbox{Half duplex, $N$ is odd;} \\
    R\left(\frac{1}{2} + \frac{1}{2N}\right), & \hbox{Half duplex, $N$ is even.} \\
\end{array}%
\right.
\end{eqnarray}
where $\tau$ is the average duration from the time a packet
arrives at the source until it reaches the destination node, and
$\doteq$ denotes asymptotic equality.
%Denote $T_m$ as the random
%number of ARQ rounds needed for the packet to be transmitted from
%node $m$ to node $m+1$, then $E(T) = \sum_{m=1}^{N-1} E(T_m)
%\doteq N-1$.
A similar argument as in \cite{GamalDamen06themimo} shows that
$E(\tau) \doteq (N-1)T$ for high SNR.

\subsection{Diversity-Multiplexing-Delay Tradeoff}

The probability of error $P_e$ in the transmission has two
sources: from the ARQ error: the packet is dropped because the
receiver fails to decode the message within the allocated number
of ARQ rounds, denoted as $P_{\mbox{\tiny{ARQ}}}$, and the
probability that a message misses its deadline at any node due to
large queueing delay, denoted as $P_{\mbox{\tiny{Queue}}}$. We
will give $P_e$ for various ARQ relay networks.
Following the framework of \cite{ZhengTse03diversity}, we assume
the size of information messages $b(\rho)$ depends on the
operating signal-to-noise ratio (SNR) $\rho$, and a family of
space time codes $\{\mathcal{C}_\rho\}$ with block rate $R(\rho) =
b(\rho)/T \triangleq r\log\rho$. We use the effective ARQ
multiplexing gain and the ARQ diversity gain
\cite{GamalDamen06themimo}
\begin{eqnarray}
r_e\triangleq \lim_{\rho \rightarrow
\infty}\frac{\eta(\rho)}{\log\rho}, \quad d \triangleq -\lim_{\rho
\rightarrow \infty} \frac{\log P_e(\rho)}{\log \rho}.
\end{eqnarray}
We cannot assume infinite SNR because otherwise the queueing delay
will be zero, as pointed out in \cite{HollidayGoldsmithPoor2008}.
However we assume high SNR to use the DMDT results in our
subsequent analysis.

\section{Diversity, multiplexing, and delay tradeoff via optimal ARQ round allocation}

\subsection{Full-Duplex Relay in Multihop Relay Network}

\subsubsection{Per-Hop Delay Constraint}

The probability of error depends on the ARQ window length
allocation $L_i$, deadline constraint $k_i$, multiplexing rate
$r$, and SNR $\rho$. For a given $r$ and $\rho$, we have
\ben && P_e(\{L_i\}, \{k_i\}|\rho, r) = \nonumber
\\
&&P_{\mbox{\tiny{ARQ}}}(\rho, \{L_i\}) +
\sum_{i=1}^{N-1}P_{\mbox{\tiny{Queue}}}(D_i>k_i). \label{cost}
\een
Here $D_i$ denotes the random delay at the $i$th link when the
queue is stationary. This $P_e$ expression is similar to that
given by Equation (33) of \cite{HollidayGoldsmithPoor2008}.
% except for the source coding distortion term (because the problem therein is joint
%source-channel coding.)
Our goal is to allocate per-hop ARQ round $\{L_i\}$ and delay
constraint $\{k_i\}$ to minimize the probability of error $P_e$.
%subject to the constraint$\sum_{i = 1}^{N-1}L_i \leq L$ and $\sum_{i = 1}^{N-1} k_i \leqk$.

For the long-term static channel, using the DMDT analysis results
\cite{YaoDenizGoldsmith2009} we have:
\ben P_{\mbox{\tiny{ARQ}}}(\rho, \{L_i\}) = \sum_{i = 1}^{N-1}
\rho^{-f_i\lsb \frac{r}{L_i} \rsb}. \een
Here $f_i(r)$ is the diversity-multiplexing tradeoff (DMT) for a
point-to-point MIMO system formed by nodes $i$ and $i+1$. Assuming
sufficient long block lengths, $f_i(r)$ is given by Theorem 2 in
\cite{ZhengTse03diversity} quoted in the following:
\begin{thm}\label{DMT}
\cite{ZhengTse03diversity} \textit{For sufficiently long block
lengths, the diversity-multiplexing tradeoff (DMT) $f(r)$ for a
MIMO system with $M_t$ transmit and $M_r$ receive antennas is
given by the piece-wise linear function connecting the points $(r,
(M_t - r)(M_r - r)),$ for $r = 0, \cdots, \min(M_t, M_r)$.}
\end{thm}

Denote the amount of time spent in the $i$th node by the $n$th
message as $D_n^i$. The probability of packet loss
$P_{\mbox{\tiny{Queue}}}(D_i>k_i)$ can be found as the limiting
distribution of $ \lim_{n\rightarrow \infty} P(D_n^i > k_i)$
(adapted from Theorem 7.4.1 of \cite{Ross1995}):
\begin{lemma}\label{lemma1}
\textit{The limiting distribution of the event that the delay at
node $i$ exceeds its deadline $k_i$, for  $M/M/1$ queue models, is
given by:}
\ben P_{\mbox{\tiny{Queue}}}(D_i >k_i) = \lim_{n \rightarrow
\infty} P(D_n^i
> k_i) = \frac{L_i}{\mu}e^{-k_i \lsb \frac{1}{L_i} - \frac{1}{\mu}
\rsb}. \een
\end{lemma}
\noindent Here the difference in the service rate and packet
arrival rate $\frac{1}{L_i} - \frac{1}{\mu} \geq 0$ and utility
factor $\frac{L_i}{\mu}$ both indicate how ``busy'' node $m$ is.
%One implication of Lemma \ref{lemma1} is that for the same $\mu$,
%the more ARQ rounds we use the larger the probability that the
%packet will miss the deadline $k_i$.
%
Using the above results, (\ref{cost}) can be written as
\ben  P_e\lsb \{L_i\}, \{k_i\} |\rho, r\rsb  =
\sum_{i=1}^{N-1}\lmb \rho^{-f_i\lsb \frac{r}{L_i} \rsb} +
\frac{L_i}{\mu} e^{-k_i\lsb \frac{1}{L_i} -
\frac{1}{\mu}\rsb}\rmb. \label{tran_dist_1}\een
Note that the queueing delay message loss error probability is
decreasing in $L_i$, and the ARQ error probability is increasing
in $L_i$. Hence an optimal ARQ rounds allocation at each node
$L_i$ should trade off these two terms. Also, the optimal ARQ
allocation should also equalize the performance of each link, as
the weakest link determines the system
performance\cite{YaoDenizGoldsmith2009}.
%These intuition guides us in designing an optimal ARQ protocol.

Hence the optimal ARQ allocation can be formulated as the
following optimization problem:
\begin{equation}\begin{split}
    %\min_{\{L_i\}, \{k_i\}}  &  D_c\lsb \{L_i\}, \{k_i\}\rsb \\
     %\min_{\{L_i\}, \{k_i\}}  &  \sum_{i=1}^{N-1}  \SNR^{-f_i\lsb
%\frac{r}{L_i} \rsb} + \frac{L_i}{\mu} e^{-k_i\lsb \frac{1}{L_i} -
%\frac{1}{\mu}\rsb}\\
    \min_{\{L_i\}, \{k_i\} \in \mathcal{A}} & P_e(\{L_i\}, \{k_i\}|\rho, r)
            \label{opt_prob}
\end{split}\end{equation}
where \begin{eqnarray}
\mathcal{A} = \left\{ \begin{array}{l}
  \sum_{i = 1}^{N-1} L_i\leq L, \\
  1 \leq L_i \leq \mu, \quad i = 1, \cdots, N-1 \\
   \sum_{i = 1}^{N-1} k_i \leq k. \\
\end{array} \right\}
            \end{eqnarray}
%
%
%The optimal solution of this problem exists under moderate
%conditions.
The following lemma (proof omitted due to the space limit)
%in Appendix \ref{app:proof_convex})
shows that the total transmission distortion function
(\ref{tran_dist}) is convex in the interior of $\mathcal{A}$.
%Hence this problem is convex and has an unique optimal solution.
%
\begin{lemma} \label{lemma_convex}
\textit{The transmission distortion  function (\ref{tran_dist}) is
convex jointly in $L_i$ and $k_i$ in the convex set}
\ben \lbb \{L_i\}, \{k_i\}: k_i>\frac{L_i}{2(\frac{\mu}{L_i}-1)},
\quad i = 1, \cdots N-1. \rbb, \nonumber \een
%
%\textit{Also, for fixed $k_m$ or $L_m$, the univariate distortion
%function in the other variable is convex.}
%
\end{lemma}
%
%
%\noindent An interpretation of the condition in Lemma
%\ref{lemma_convex} is that: a large the ratio of the inter-arrival
%time and the service time $\frac{\mu}{L_i}$ and a small the
%service time $L_i$ means a smaller expected delay on this node,
%and this also allows a smaller $k_i$.
Lemma \ref{lemma_convex} says that except for the ``corners'' of
$\mathcal{A}$ the cost function is convex. However these
``corners'' have higher probability of error: $k_i$ and $L_i$ take
extreme values and hence one link may have a longer queueing delay
then the others.
%Hence the optimal solution is less likely
%to lie in the corners and
So we only need to search the interior of $\mathcal{A}$ where the
cost function is convex.

%\subsubsection{A Marginal Cost Interpretation}

To gain some insights into where the optimal solution resides in
the feasible domain for the above problem, we present a marginal
cost interpretation. Note that the probability of error can be
decomposed as a sum of probability of error on the $i$th link. The
optimal ARQ rounds allocated on this link should equalize the
``marginal cost'' of the ARQ error and the packet loss due to
queueing delay. For node $i$, with fixed $k_i$, the marginal costs
(partial differentials) of the ARQ error probability, and the
packet loss probability due to queueing delay, with respect to
$L_i$ are given by
\ben
\frac{\partial \rho^{-f_i\lsb \frac{r}{L_i} \rsb}}{\partial L_i} =
\frac{r}{L_i^2} f'_i\lsb \frac{r}{L_i} \rsb \rho^{-f_i\lsb
\frac{r}{L_i}\rsb} \ln \rho < 0, \een
and \ben
\frac{\partial P_{\mbox{\tiny{Queue}}}(D_i>k_i)}{\partial L_i} =
\frac{1}{\mu} \lsb 1+\frac{k}{L_i} \rsb e^{-k_i \lsb \frac{1}{L_i}
-\frac{1}{\mu} \rsb}
> 0. \een
Note that $f'_i < 0$. The optimal solution equalizes these two
marginal costs by choosing $L_i \in [1, \mu]$. Note that these
marginal cost functions are monotone in $L_i$, hence the
equalizing $L_i^*$ exists and $1< L_i^*< \mu$ if the following two
conditions are true for $L_i = 1$ and $L_i = \mu$:
\ben (i): \left. \frac{\partial
P_{\mbox{\tiny{Queue}}}(D_i>k_i)}{\partial L_i} \right|_{L = 1} <
-\left.\frac{\partial \rho^{-f_i\lsb \frac{r}{L_i} \rsb}}{\partial L_i} \right|_{L = 1},\\
(ii): \left. \frac{\partial
P_{\mbox{\tiny{Queue}}}(D_i>k_i)}{\partial L_i} \right|_{L = \mu}
> -\left.\frac{\partial \rho^{-f_i\lsb \frac{r}{L_i}
\rsb}}{\partial L_i} \right|_{L = \mu}, \label{cond_2} \een
These conditions involve nonlinear inequalities involving $\mu$,
$\rho$, $r$, $M_i$ and $M_{i+1}$, which defines the case when the
optimal solution is in the interior of $\mathcal{A}$. Analyzing
these conditions reveals that these conditions tend to satisfy at
lower multiplexing gain $r$, small $M_i$ or $M_{i+1}$, small
$k_i$, and larger $\mu$ (light traffic). Note that with high SNR
condition (ii) is always true for moderate $k$ values. When $(i)$
and $(ii)$ are violated, which means one error dominates the
other, then the optimal solution lies at the boundary of
$\mathcal{A}$.
%Hence If
%condition
%when $k_i$ is very large and (i) is violated, the optimal solution
%will be pushed to the maximum ARQ rounds allowed.
%
With the total ARQ rounds constraint in (\ref{opt_prob}), using
the Lagrangian multiplier an argument similar to above still
holds.

\subsubsection{End-to-End Delay constraint}

When the buffer per node is large enough a per hop delay
constraint is not needed, and we can instead impose an end-to-end
delay constraint. The exact expression for the tail probability of
the end-to-end delay is intractable. However a large deviation
result is available. The following theorem can be derived using
the main theorem in \cite{Ganesh1998}:
\begin{thm}\label{SojournTimeFullDuplex}
\textit{For a stationary $M/M/1$ queue tandem (with full-duplex
relays):}
\begin{eqnarray}
\lim_{k\rightarrow \infty} \lim_{n \rightarrow
\infty}\frac{1}{k}\log
P_{\mbox{\tiny{Queue}}}\left(\sum_{i=1}^{N-1} D_n^i \geq k\right)
= -\theta^*, \nonumber
\end{eqnarray}
\textit{where} $\theta^* = \min_{i=1}^{N-1}
\left\{\frac{1}{L_i}-\frac{1}{\mu}\right\}$.
\end{thm}
\vspace{0.1in}
This theorem says that the bottleneck of the queueing network is
the link with longest mean service time $L_i$. Hence the optimal
ARQ round allocation problem can be formulated as:
\begin{eqnarray}
\min_{\{L_i\} \in \mathcal{B}} & P_e(\{L_i\}, \{k_i\}|\rho, r)
            \label{opt_prob_e2e}
\end{eqnarray}
where
\ben  &&P_e\lsb \{L_i\}, \{k_i\} |\rho, r\rsb  \nonumber \\
&&=P_{\mbox{\tiny{ARQ}}}(\rho, \{L_i\}) +
P_{\mbox{\tiny{Queue}}}\left(\sum_{i=1}^{N-1} D_n^i \geq k\right),
\nonumber
\\
&& \doteq \sum_{i=1}^{N-1} \rho^{-f_i\lsb \frac{r}{L_i} \rsb} +
e^{-\theta^* k}. \label{tran_dist}\een
\begin{eqnarray} \mathcal{B} = \left\{
\begin{array}{l}
  \sum_{i = 1}^{N-1} L_i\leq L, \\
  1 \leq L_i \leq \mu, \quad i = 1, \cdots, N-1
\end{array} \right\}
            \end{eqnarray}
For high SNR, this can be shown to be a convex optimization
problem.
%
%By allocating the per-hop delay constraint meets the end-to-end
%delay constraint though it may not be the optimal one.
%
%
A simple argument can show that the packet loss probability with
the per-hop delay constraint is larger than that using the more
flexible end-to-end constraint.
%, for which the packet can delay
%more than
%$k_i$ as long as the end-to-end delay is smaller than $k$.
%However, designing this
%flexible end-to-end strategy requires stationary distribution of
%waiting time vector for single server queues in series, which is
%generally intractable except special cases using the heavy traffic
%approximation (see, e.g., \cite{Harrison1973}). Our proposed
%strategy is close to optimal (we optimize over the $k_i$s) yet
%with tractable solution.

\subsection{Half-duplex Relay in Multihop Network}

Half-duplex relay is not a standard queue tandem model. However we
can also derive a large deviation result for the tail probability
for the end-to-end delay of a multihop network with half-duplex
relays (proof in the Appendix):
\begin{thm}\label{SojournTimeHalfDuplex}
\textit{For a stationary $M/M/1$ queue tandem (with half-duplex
relays), when the number of node $N$ is large:}
\begin{eqnarray}
\lim_{k\rightarrow \infty} \lim_{n \rightarrow
\infty}\frac{1}{k}\log
P_{\mbox{\tiny{Queue}}}\left(\sum_{i=1}^{N-2} D_n^i \geq k\right)
= -\theta^*.
\end{eqnarray}
%\textit{where } $\theta^* = \min_{m=1}^{N-1}
%\left\{\frac{1}{L_m}-\frac{1}{\mu}\right\}$.
\end{thm}
From this theorem we conclude that the optimal ARQ allocation
problem with the end-to-end constraint and half-duplex relays can
be formulated the same as that with full-duplex relays (\ref{opt_prob_e2e}). % (a stretched version of the DMDT for the full-duplex
%relay case.)

\section{Numerical Examples} \label{egs}

Consider a MIMO relay network consists of a source, a relay, and a
destination node. The relay is full-duplex. The number of antennas
on each node is $(M_1, M_2, M_3)$, $M_1 = 4$, $M_2 = 1$, and $M_3
= 2$, where the relay has a single antenna. Other parameters are:
$\rho = 20$dB, $k = 30$, $L = 8$, and the multiplexing gain is $r
= 2$. The base 10 logarithm of the cost function
(\ref{tran_dist_1}) is shown in Fig. \ref{Fig:3Deg}. We have
optimized the cost function with respect to $L_2$ and $k_2$ so we
can display it in three dimensions. Note that the surface is
convex in the interior of the feasible region. The optimal
$L_1^*$, $L_2^*$, $k_1^*$ are shown in Fig. \ref{Fig:threeHop422}.
Also note that as $r$ increases to the maximum possible $r = 4$,
the total number of ARQ rounds allocated $L_1^*+L_2^*$ gradually
increases to the upper bound $L = 8$ as $k$ increases.

%\begin{figure}[h]
%    \begin{center}
%        \includegraphics[width=4in]{opt_L1}
%        \includegraphics[width=4in]{opt_k1}
%    \end{center}
%    \caption{Allocations of optimal ARQ versus the fixed ARQ,
%    in a (4, 1, 2) MIMO relay network. SNR is 20 dB. (The optimal $k_2^* = k - k_1^*$.)}
%    \label{Fig:threeHop}
%\end{figure}

%\begin{figure}[h]
%    \begin{center}
%        \includegraphics[width=4in]{cost}
%    \end{center}
%    \caption{Total transmission distortion of
%    the optimal ARQ versus the fixed ARQ, in a (4, 1, 2) MIMO relay network. SNR is 20 dB.}
%    \label{Figcost}
%\end{figure}

\section{Conclusions and Future Work}

We have studied the diversity-multiplexing-delay tradeoff in
multihop MIMO networks by considering an optimal ARQ allocation
problem to minimize the probability of error, which consists of
the ARQ error and the packet loss due to queueing delay. Our
contribution is two-fold: we combine the DMDT analysis with
queueing network theory, and we use the tail probability of random
delay to find the probability of packet loss due to queueing
delay. Numerical results show that optimal ARQ should equalize the
performance of each link and avoid long service times that cause
large queueing delay. Future work will investigate joint
source-channel coding in multihop MIMO relay networks, extending
the results of \cite{HollidayGoldsmithPoor2008}.

%Our current approach assumes stable queueing model. An interesting
%future topic would be study the dynamic optimal ARQ allocation
%based on instantaneous queueing status using the dynamic
%optimization tool. This would be an extension of dynamic ARQ
%allocation proposed in \cite{HollidayGoldsmithPoor2008} from the
%point-to-point MIMO case to the multihop MIMO relay networks.

\appendix

\noindent \textbf{Proof of Theorem
\ref{SojournTimeHalfDuplex}}\label{app:half_duplex}

For node $i$, $i = 1\cdots N$, let the random variable $S_n^i$
denotes the service time required by the $n$th customer at the
$i$th node (the number of ARQs used for the $n$th packet), and
$A_n^i$ be the inter arrival time of the $n$th packets (i.e., the
time between the arrival of the $n$th and $(n-1)$th packages to
this node). The waiting time of the $n$th packet at the $i$th node
$W_n^i$ satisfies Lindley's recursion (see \cite{Ganesh1998}):
\ben W_n^i = (W_{n-1}^i + S_{n-1}^{i+1} - A_{n}^i)^+, \quad 2\leq
i\leq N-2, \label{rec_2}\een
where $(x)^+ = \max(x, 0)$.  The total time a message spent in a
node is its waiting time plus its own service time, hence
\begin{eqnarray} D_n^i = W_n^i + S_n^i.
\end{eqnarray}
The arrival process to the $(i+1)$th
node is the departure process from the $i$th node, which satisfies
the recursion:
\begin{eqnarray}
A_n^i = A_n^{i-1} + D_n^{i-1} - D_{n-1}^i,  \quad 2\leq i\leq N-2.
\label{rec_1}
\end{eqnarray}
with $A_n^i$ a Poisson process with rate $1/\mu$. Also the waiting
time at the source satisfies:
\begin{eqnarray}
W_n^1 = (W_{n-1}^1 + S_{n-1}^1 + S_{n-1}^2 - A_n^1)^+.
\end{eqnarray}
A well-known result is that (see, e.g. \cite{Ganesh1998}), if the
arrival and service processes satisfy the stability condition,
then the Lindley's recursion has the solution:
\begin{eqnarray}
W_n^i &=& \max_{j_i\leq n}(\sigma^i_{j_i, n-1} -
\tau^i_{j_i+1,n}),
\quad i = 2, \cdots N-2, \nonumber \\
W_n^1 &=& \max_{j_1\leq j_2} (\sigma^1_{j_1, j_2 -1} +
\sigma^2_{j_1, j_2-1} - \tau^1_{j_1+1, j_2}).
\end{eqnarray}
where the partial sum $\tau_{l,p}^i = \sum_{k=l}^p A_k^i$ and
$\sigma_{l,p} = \sum_{k=l}^p S_k^i$. Hence
\begin{eqnarray}
D_n^i = \max_{j_i\leq n}(\sigma^i_{j_i, n-1} + S_n^i -
\tau^i_{j_i+1,n}), \quad i = 2, \cdots N-2. \label{D}
\end{eqnarray}
From (\ref{rec_1}) we have $\tau_{l,p}^i = \tau_{l,p}^{i-1} +
D_p^{i-1} - D_{l-1}^{i-1}$ for $l\leq p+1$, and 0 otherwise. Plug
this into (\ref{D}) we have
\begin{eqnarray}
D_n^i = \max_{j_i\leq n}(\sigma^{i+1}_{j_i, n-1} + S_n^i -
\tau_{j_i+1,n}^{i-1} - D_n^{i-1} + D_{j_i}^{i-1}).
\end{eqnarray}
Hence the recursive relation if we move $D_n^{i-1}$ to the
left-hand-side:
\begin{eqnarray}
D_n^i + D_n^{i-1} = \max_{j_i\leq n}(\sigma_{j_i, n-1}^{i+1} +
S_n^i -\tau_{j_i+1,n}^{i-1} + D_{j_i}^{i-1}). \label{seq}
\end{eqnarray}
Now from (\ref{D}) we have $D_{j_i}^{i-1} = \max_{j_{i-1}\leq
j_i}(\sigma^i_{j_{(i-1)},j_i-1} + S_{j_i}^{i-1} -
\tau_{j_{(m-1)}+1, j_m}^{i-1})$. Plug this in the above
(\ref{seq}) we have
\begin{eqnarray}
&& D_n^i + D_n^{i-1}\nonumber\\
% &=& \max_{j_{(i-1)}\leq j_i \leq n}
%(\sigma_{j_i, n-1}^{i+1} + S_n^i -\tau_{j_i+1,n}^{i-1} +
%\sigma^i_{j_{(i-1)},j_i-1} + S_{j_i}^{i-1} - \tau_{j_{(i-1)}+1,
%j_i}^{i-1}) \nonumber \\
&=& \max_{j_{(i-1)}\leq j_i \leq n} (\sigma_{j_i, n-1}^{i+1} +
S_n^i  + \sigma^i_{j_{(i-1)},j_i-1} + S_{j_i}^{i-1} -
\tau_{j_{(i-1)}+1, n}^{i-1}) \nonumber
\end{eqnarray}
Do this inductively, we have
\begin{eqnarray}
\sum_{i=2}^{N-2} D_n^i = \max_{j_2\leq \cdots \leq
j_{N-1}=n}\left[ \sum_{m=2}^{N-2}(\sigma^{i+1}_{j_i,j_{(i+1)}-1} +
S_{j_{i+1}}^i) -\tau^1_{j_2+1, n} \right]. \nonumber
\end{eqnarray}
If we also add $D_n^1 = W_n^1 + S_n^1$ to the above equation,
after rearranging terms we have:
\begin{eqnarray}
&&\sum_{i=1}^{N-2} D_n^i =
\sum_{i=2}^{N-2}(\sigma^i_{j_{(i-1)},j_i-1} +
S_{j_{(i+1)}}^i)-\tau^1_{j_2+1, n}\nonumber \\
&& + S_{j_2}^1 + \sigma_{j_1, j_2-1}^1 + S_{j_2}^2 +
\sigma_{j_{N-2},j_{(N-1)}-1}^{N-1}. \label{final}
\end{eqnarray}
Note that $\sigma^i_{j_{(i-1)},j_i-1}$ is independent of
$S_{j_{(i+1)}}^i$. For long queue we can ignored the last four
terms caused by edge effect (the source and end queue of the
multihop relay network). By stationarity of the service process
$\sigma^i_{j_{(i-1)},j_i-1} + S_{j_{(i+1)}}^i$ has the same
distribution as $\sigma^i_{0, j_i-j_{(i-1)}}$. Then (\ref{final})
reduces to the case studied in \cite{Ganesh1998} and we can borrow
the large deviation argument therein to derive the exponent
$\theta^*$.

\vspace{0.3in}

\bibliography{yao_proposal_2}

\end{document}